\documentclass[aps,prl,superscriptaddress,showpacs,showkeys,reprint]{revtex4-1}
\usepackage{amsmath}
\usepackage{amsfonts}
\usepackage{amsthm}
\usepackage{graphics}
\usepackage{graphicx}
\usepackage{subfigure}
\usepackage{amssymb}
\usepackage{color}
\usepackage{url}
\usepackage[abs]{overpic}
\usepackage{tikz}
\usepackage{bm}
\usepackage[normalem]{ulem}

\definecolor{darktangerine}{rgb}{1.0, 0.66, 0.07}

\begin{document}

\title{Sticking Transition in a Minimal Model for the Collisions of Active Particles in
Quantum Fluids}

\author{Vishwanath Shukla}
\email{research.vishwanath@gmail.com}
\affiliation{Laboratoire de Physique Statistique de l'Ecole Normale 
Sup{\'e}rieure, 24 Rue Lhomond, 75231 Paris, France}
\affiliation{Centre for Condensed Matter Theory, Department of Physics, 
Indian Institute of Science, Bangalore 560012, India.}
\author{Marc Brachet}
\email{brachet@physique.ens.fr}
\affiliation{Laboratoire de Physique Statistique de l'Ecole Normale 
Sup{\'e}rieure, \\
associ{\'e} au CNRS et aux Universit{\'e}s Paris VI et VII,
24 Rue Lhomond, 75231 Paris, France}
\author{Rahul Pandit}
\email{rahul@physics.iisc.ernet.in}
\altaffiliation[\\ also at~]{Jawaharlal Nehru Centre For Advanced
Scientific Research, Jakkur, Bangalore, India.}
\affiliation{Centre for Condensed Matter Theory, Department of Physics, 
Indian Institute of Science, Bangalore 560012, India.} 

\date{\today}
\begin{abstract}

Particles of low velocity, travelling without dissipation in a superfluid, can
interact and emit sound when they collide. We propose a minimal model in which
the equations of motion of the particles, including a short-range repulsive
force, are self-consistently coupled with the Gross-Pitaevskii equation. We use
this model to demonstrate the existence of an effective superfluid-mediated
attractive interaction between the particles; and we study numerically the
collisional dynamics of particles as a function of their incident kinetic energy
and the length-scale of the repulsive force. We find a transition 
from almost elastic to completely inelastic (sticking) collisions as the
parameters are tuned. We find that aggregation and clustering result from this sticking transition in
multi-particle systems.

\end{abstract}
\maketitle

Studies  
of an assembly of particles in a superfluid
have a rich history~\cite{donnellybook}. This challenging problem is of relevance
to recent experiments on particles in superfluid
helium~\cite{bewley2006superfluid,bewley2008characterization,
bewleyparticlesdetails,
He2excimerT0tracer,mantiaprbaccpdf}
and impurities in cold-atom Bose-Einstein condensates
(BECs)~\cite{Spethmann2012}. Its understanding requires models and
techniques from the physics of quantum fluids with
state-of-the-art methods from theoretical and numerical studies
of turbulence. In contrast to particles moving through a viscous
fluid, particles move through a zero-temperature superfluid
without dissipation, so long as they travel at speeds lower than
the critical speed above which the particles shed quantum
vortices~\cite{frisch1992vcrit,nore2000subcritdissp,winiecki2}.
The motion of a single particle, which is affected by the
superflow and acts on it too, has been studied in
Ref.~\cite{winiecki1} in a Gross-Pitaevskii (GP) superfluid. We
refer to this as an \textit{active particle}.

We go well beyond earlier
studies~\cite{winiecki1,pitapart04,Varga2015} of this problem by developing
a minimal model. 
We introduce
\textit{active and interacting} particles in the Gross-Pitaevskii
Lagrangian that describes a weakly interacting superfluid at zero
temperature.  By
using this model we show that, even if particles move through the
superfluid at subcritical speeds, they can dissipate energy when
they collide, because a two-particle collision excites sound
waves; clearly the coefficient of restitution
$e < 1$, for such a collision. Furthermore, we demonstrate that
there is a superfluid-mediated attraction between the
particles.  We calculate this attraction both approximately, via
a Thomas-Fermi approximation, and numerically, from a direct
numerical simulation (DNS) of the Gross-Pitaevskii Equation (GPE).  We
show that the interplay between the short-range (SR) particle
repulsion, which we have included in our Lagrangian, and the
superfluid-mediated (SM) attraction leads to a \textit{sticking} transition at
which the coefficient of restitution $e$ for two-particle collisions vanishes. We develop a
simple, mean-field theory for this transition and we compare it
with our DNS results.  Furthermore, we elucidate the rich
dynamical behaviors of (a) two-particle collisions in the
superfluid, when the impact parameter $b$ is nonzero, and (b)
assemblies of particles, which aggregate because of the SM attraction.

To study the dynamics of particles in a Bose superfluid, we
propose the Lagrangian
\begin{equation}\label{eq:Lagrangianfull}
\begin{split}
	\mathcal{L} &= \int_{\mathcal{A}}\Bigl[\frac{i\hbar}{2}\Bigl(\psi^*\frac{\partial \psi}{\partial t}
-\psi\frac{\partial \psi^*}{\partial t}\Bigr) - \frac{\hbar^2}{2m}\nabla\psi\cdot\nabla\psi^*
+\mu|\psi|^2 \\ 
&- \frac{g}{2}|\psi|^4
-\sum^{\mathcal{N}_0}_{i=1}V_{\mathcal{P}}(\mathbf{r}-\mathbf{q}_i)|\psi|^2\Bigr]d\mathbf{r}
+\frac{m_{o}}{2}\sum^{\mathcal{N}_0}_{i=1}\dot{q}^2_{i}\\
&-\sum^{\mathcal{N}_0,\mathcal{N}_0}_{i,j,i\neq j}\frac{\Delta_E r^{12}_{SR}}{|\mathbf{q}_i-\mathbf{q}_j|^{12}},
\end{split}
\end{equation}
where $\psi$ is the complex, condensate wave function, 
$\psi^*$ its complex conjugate,
${\mathcal{A}}$
the simulation domain, $g$ the effective interaction strength,
$m$ the mass of the bosons, $\mu$ the chemical potential,
$V_{\mathcal{P}}$ the potential that we use to represent the
particles, and $\mathcal{N}_0$ the total number of particles with
mass $m_{o}$.  The last term in Eq.~(\ref{eq:Lagrangianfull}) is
the SR repulsive, two-particle potential; we treat $\Delta_E$ and
$r_{SR}$ as parameters.

\begin{figure*}
\centering
\resizebox{\linewidth}{!}{
\includegraphics[width=0.9\linewidth]{1a.eps}
\put(-340.,315){\huge{\bf (a)}}
\put(-405,345){\small{\bf (i)}}
\put(-375,315){\small{\bf (ii)}}
\put(-370,235){\small{\bf (iii)}}
\put(-355,290){\small{\bf (iv)}}
\put(-330,265){\small{\bf (v)}}
\put(-325,240){\small{\bf (vi)}}
\put(-250,262){
\includegraphics[width=0.15\linewidth]{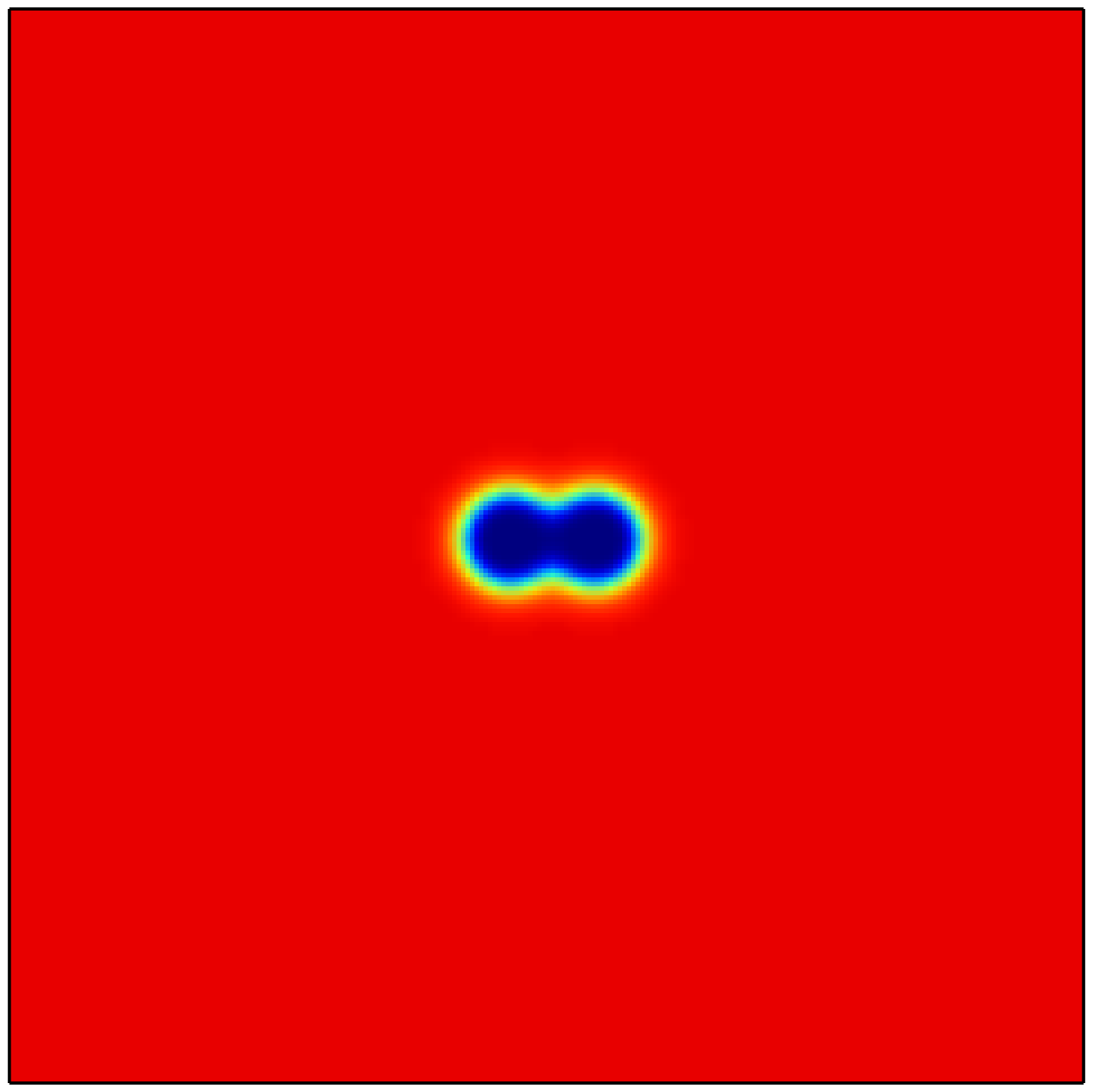}
\put(-70,60){\small{\bf (i)}}}
\put(-175,262){
\includegraphics[width=0.15\linewidth]{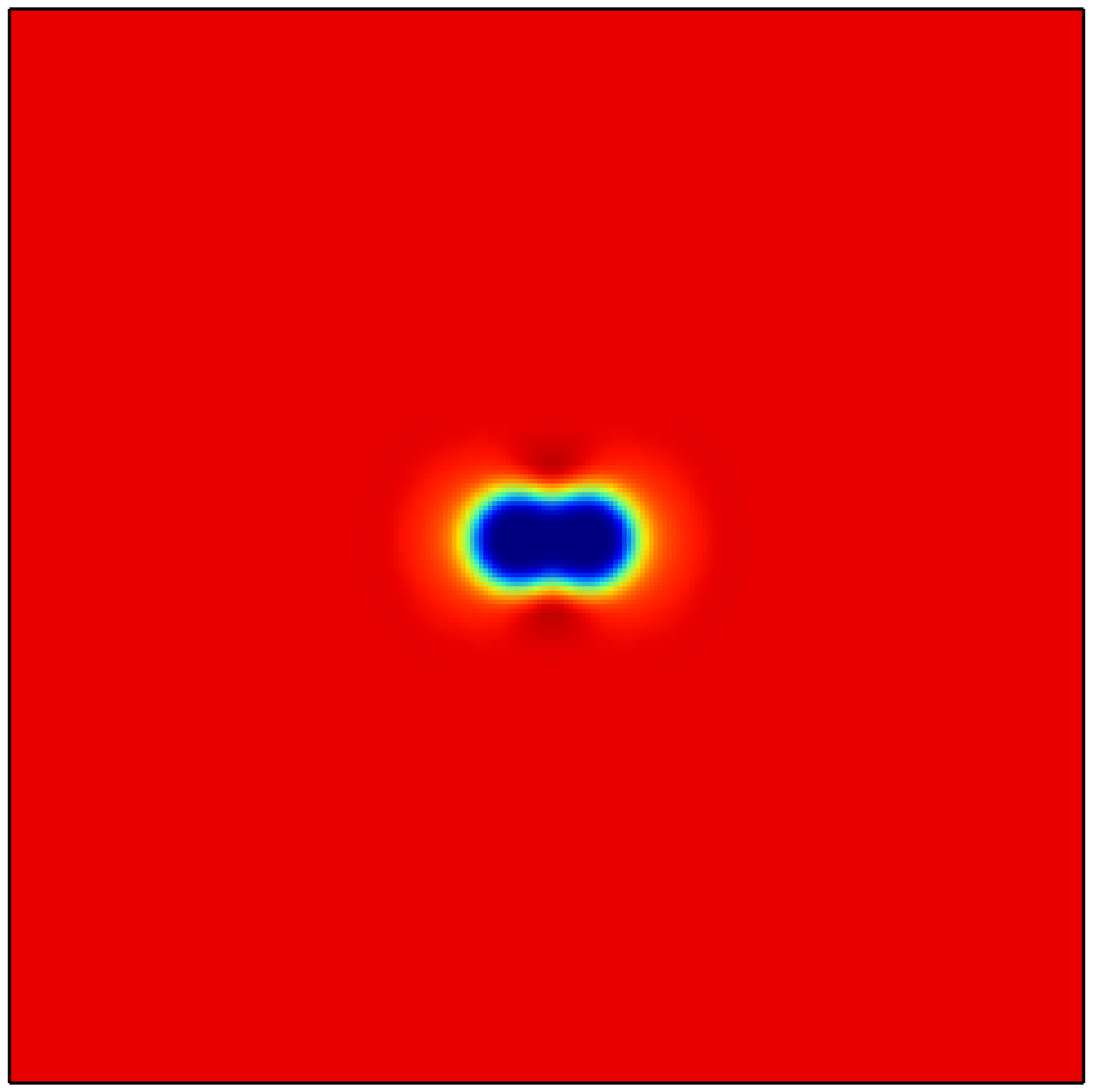}
\put(-70,60){\small{\bf (ii)}}}
\put(-100,262){
\includegraphics[width=0.15\linewidth]{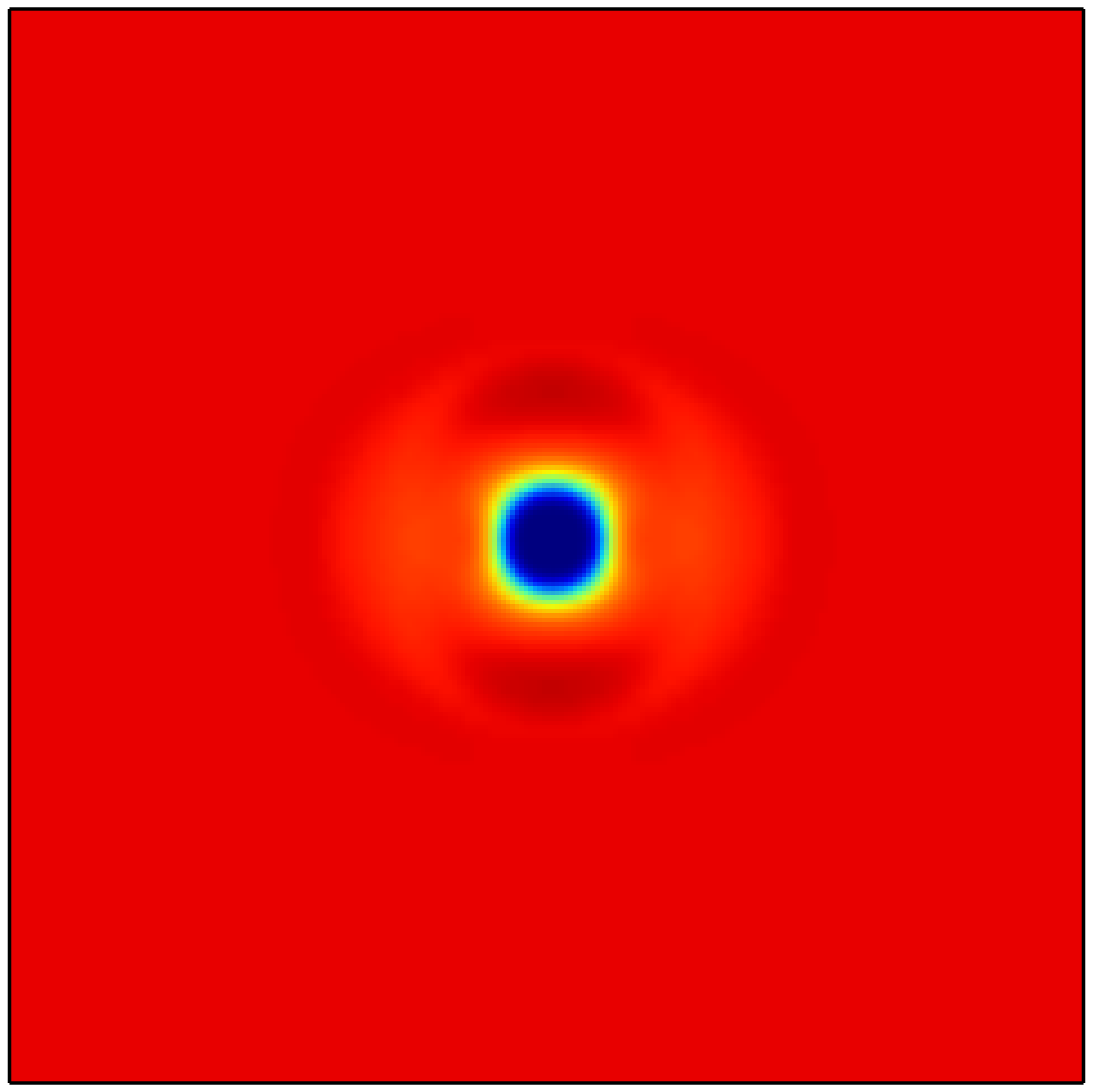}
\put(-70,60){\small{\bf (iii)}}}
\put(-250,52){
\includegraphics[width=0.15\linewidth]{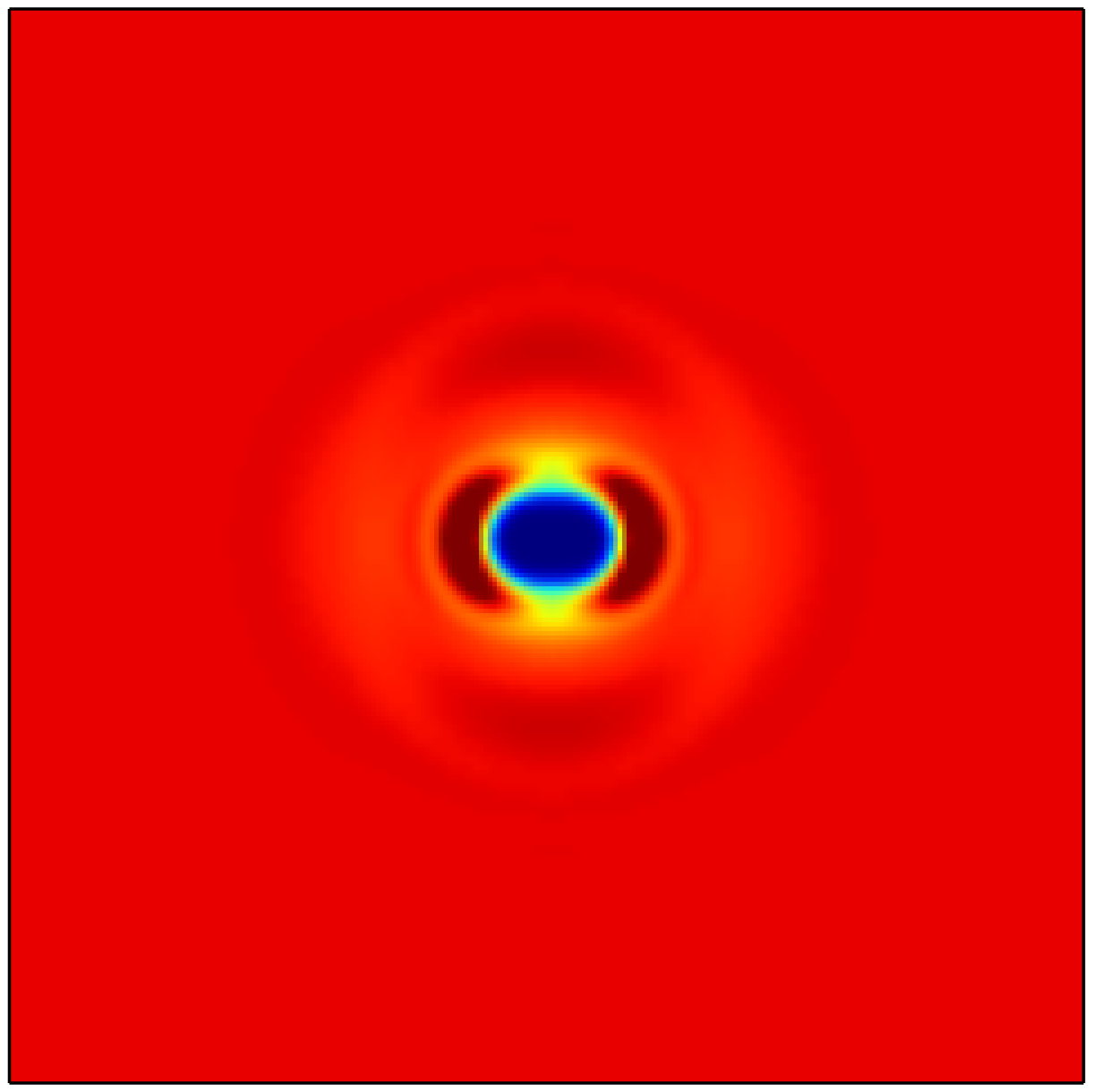}
\put(-70,60){\small{\bf (iv)}}}
\put(-175,52){
\includegraphics[width=0.15\linewidth]{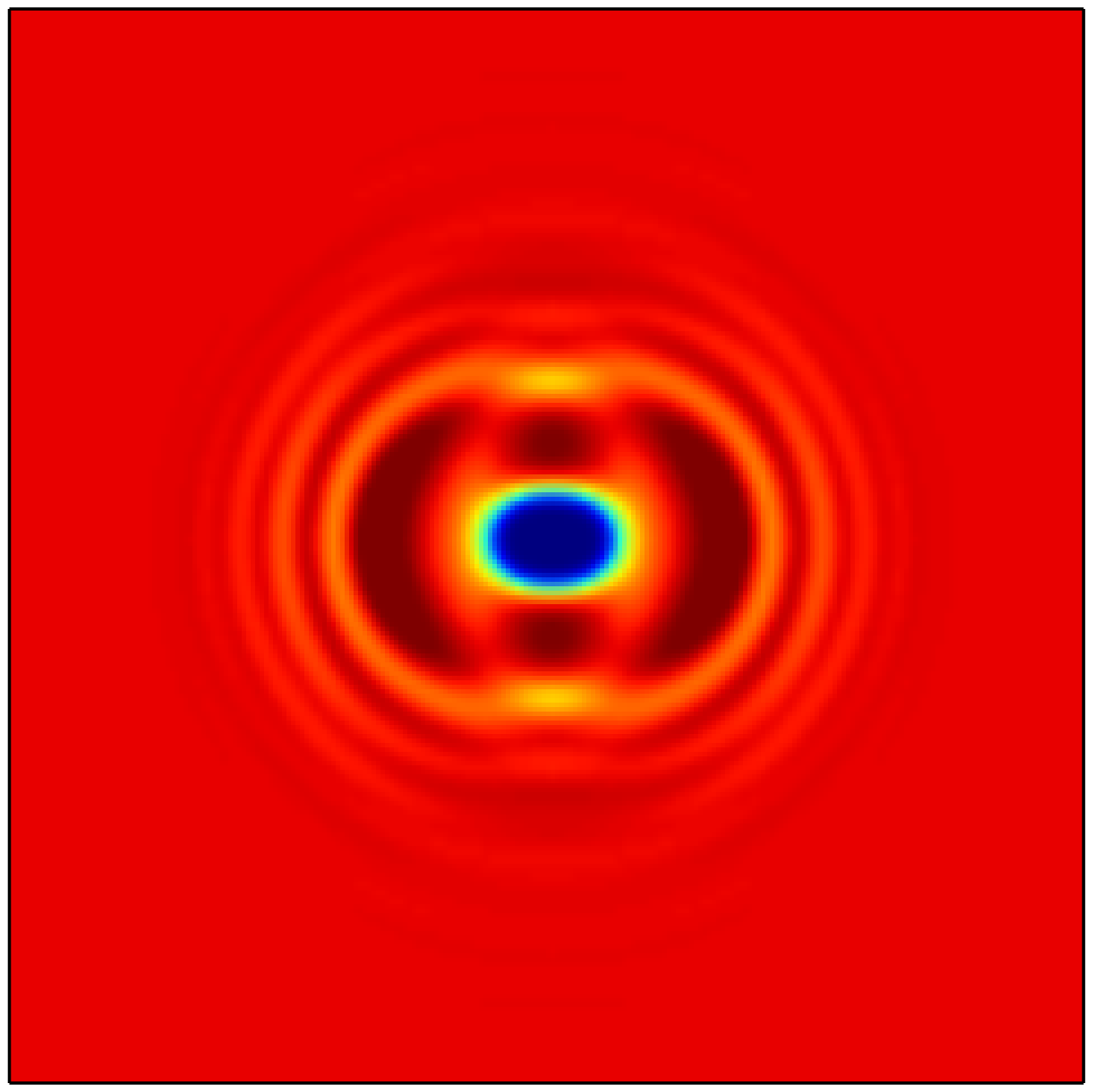}
\put(-70,60){\small{\bf (v)}}}
\put(-100,52){
\includegraphics[width=0.15\linewidth]{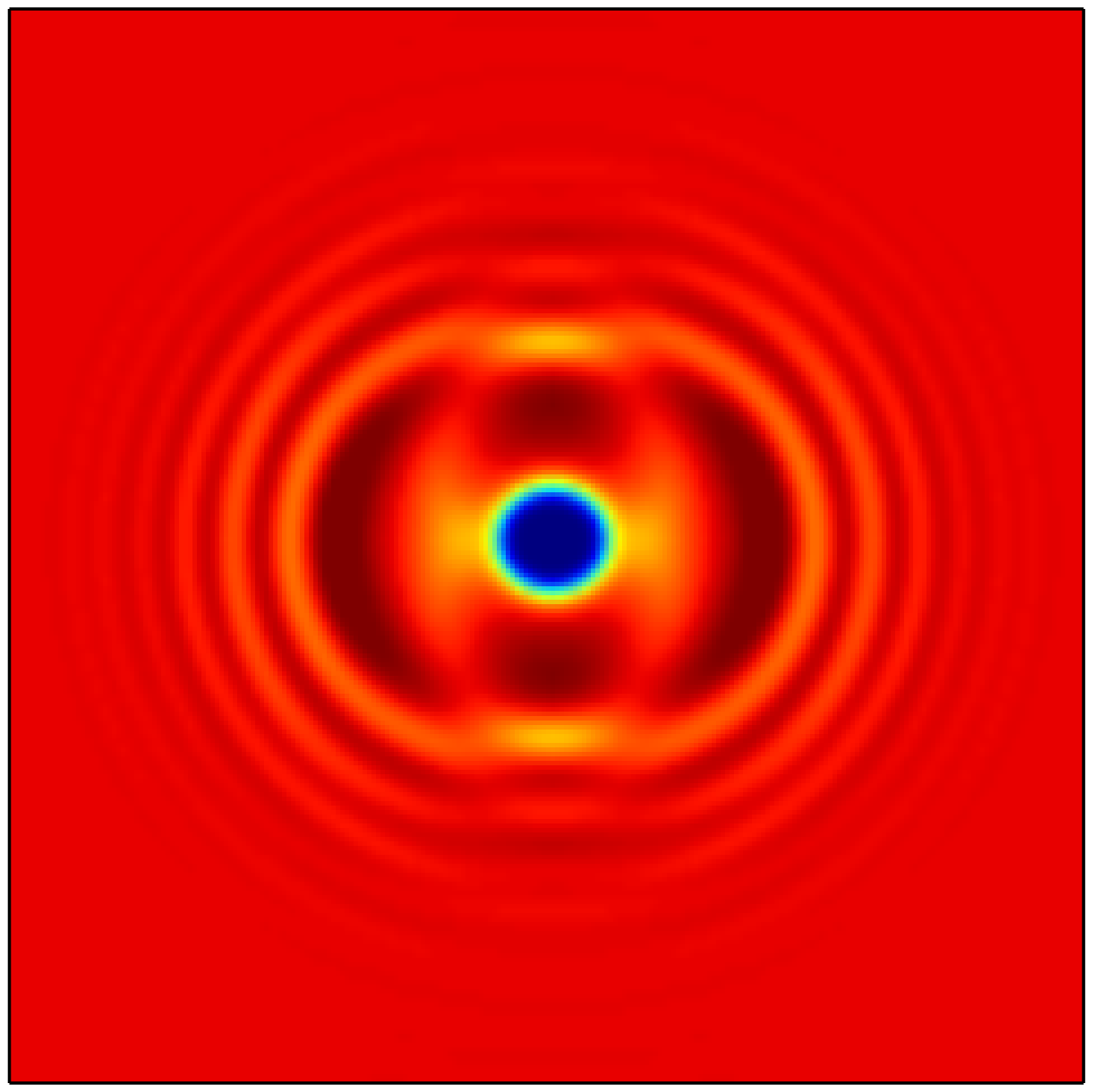}
\put(-70,60){\small{\bf (vi)}}}
\includegraphics[width=0.9\linewidth]{1b.eps}
\put(-100.,85){\small{\bf (a)}}
\put(-280,50){
\includegraphics[width=0.48\linewidth]{1b1.eps}}
\put(-320.,315){\huge{\bf (b)}}
}
\caption{\small (Color online) {\bf Superfluid-mediated attractive potential:}
(a) Plot of the particle ($\mathcal{M}=1$) positions $q_{o,x}$ versus the
scaled time $ct/\xi$. 
Inset: the sequence of the collision events shown via
the pseudocolor plots of the density field $\rho(\mathbf{r})$ 
(the particles appear as blue disks in which $\rho=0$);
particles are released from rest, with an initial
separation $r_0=7\,\xi$, they undergo multiple collisions 
with the generation of sound waves in the wake of this collision;
and they form a bound state with $r \simeq r_{\rm SR}$. 
(b) Plot of the superfluid-mediated attractive potential $U_{\rm A}$
versus the separation between the particles $r/\xi$ obtained from our DNSs;
the inset shows the same plot, but evaluated by using the Thomas-Fermi approximation
Eq.~(\ref{eq:potTF}).
}
\label{fig:fig1}
\end{figure*}

The Lagrangian (\ref{eq:Lagrangianfull}) yields the GPE
\begin{equation}\label{eq:GPEfull}
i\hbar\frac{\partial \psi}{\partial t} = -\frac{\hbar^2}{2m}\nabla^2\psi -\mu\psi + g|\psi|^2\psi
+ \sum^{\mathcal{N}_0}_{i=1}V_{\mathcal{P}}(\mathbf{r}-\mathbf{q}_i)\psi;
\end{equation}
and the equation of motion for the particle $i$
\begin{equation}\label{eq:eqmpartfull}
	m_{o}\ddot{\mathbf{q}}_i = \mathbf{f}_{o,i} + \mathbf{f}_{SR,i},
\end{equation}
where
\begin{equation}
	\mathbf{f}_{o,i} = \int_{\mathcal{A}}|\psi|^2\nabla V_{\mathcal{P}}d\mathbf{r};
\end{equation}
$\mathbf{f}_{SR,i}$ arises from the SR repulsive potential (the
last term in Eq.~(\ref{eq:Lagrangianfull})). 
In the
absence of any external force, the total energy of this system is
conserved. Moreover, the dynamical evolution of the coupled set
of Eqs.~(\ref{eq:GPEfull})-(\ref{eq:eqmpartfull}) conserves the
total momentum and the number of bosons, which constitute
the superfluid.  We can express the GP in
terms of hydrodynamical variables by using the Madelung
transformation
$\psi(\mathbf{r},t)=\sqrt{\rho(\mathbf{r},t)/m}\exp(i\phi(\mathbf{r},t))$,
where $\rho(\mathbf{r},t)$ and $\phi(\mathbf{r},t)$ are the
density and phase fields, respectively; the superfluid velocity
is $\mathbf{v}(\mathbf{r},t)=(\hbar/m)\nabla\phi(\mathbf{r},t)$,
which shows that the motion is irrotational in the absence of any
vortices.  We represent a particle by the
Gaussian potential
$V_{\mathcal{P}} = V_o\exp(-r^2/2d^2_p)$;
here $V_o$ and $d_{\rm p}$ are the strength of
the potential and its width, respectively. The
particle displaces some
superfluid with an area of the order of the particle area; we denote the
mass of the displaced superfluid by $m_f$. We use the ratio
$\mathcal{M}\equiv m_o/m_f$ to define three types of particles:
(1) heavy ($\mathcal{M}>1$), (2) neutral ($\mathcal{M}=1$), and
(3) light ($\mathcal{M}<1$).

To solve Eq.~(\ref{eq:GPEfull})-(\ref{eq:eqmpartfull}) numerically, we use a pseudospectral
method with the $2/3$-dealiasing rule~\cite{Got-Ors,vmrnjp13}, on a 2D, periodic, computational domain of
side $L=2\pi$, i.e., $\mathcal{A}=L^2$; we use a fourth-order, Runge-Kutta
scheme for time marching.  We work with the quantum of circulation
$\kappa\equiv h/m\equiv4\pi\alpha_{\rm 0}$, speed of sound $c=\sqrt{2\alpha_{\rm
0}g\rho_{\rm 0}}$, and healing length $\xi=\sqrt{\alpha_{\rm 0}/(g\rho_{\rm
0})}$. In all our calculations, we set $\rho_{\rm 0}=1$, $c=1$, and $\xi=1.44\,
dx$, where $dx=L/N_c$, $N_c=128$ is the number of collocation points, $\mu=g$,
$V_{\rm o}=10\, g$, $d_{\rm p}=1.5\, \xi$, and $\Delta_{\rm E}=0.062$.
[See the Supplemental Material~\cite{suppmat} for details.]
\begin{figure*}
\centering
\resizebox{\linewidth}{!}{
\includegraphics[width=0.3\linewidth]{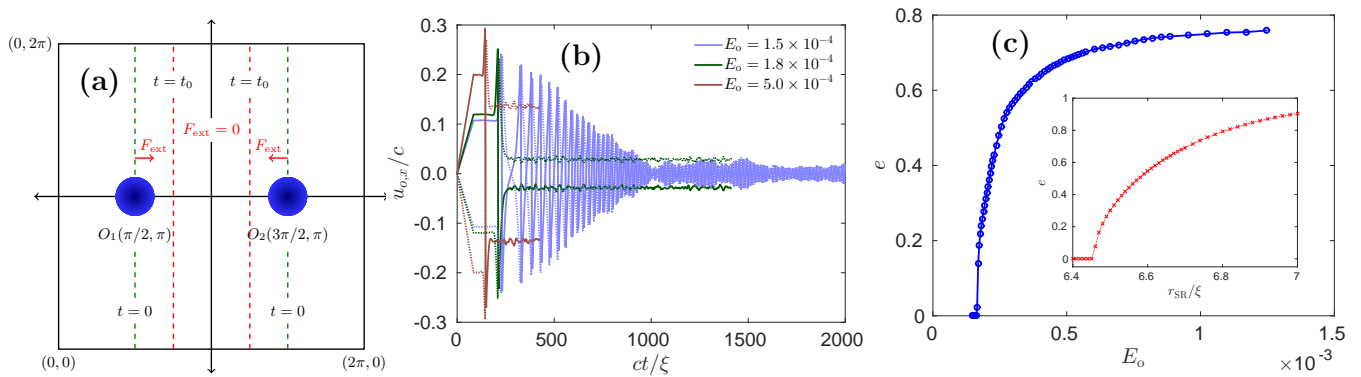}
\put(-125.,115){\large{\bf (a)}}
\includegraphics[width=0.38\linewidth]{2b.eps}
\put(-125.,125){\large{\bf (b)}}
\includegraphics[width=0.38\linewidth]{2c.eps}
\put(-145.,130){\large{\bf (c)}}
\put(-130,30){
\includegraphics[width=0.21\linewidth]{2c1.eps}}
}
\caption{\small (Color online) {\bf Head-on collisions:}
(a) Schematic diagram outlines the initial configuration and 
the procedure that we use to study
the head-on collision between two particles (blue disks).
(b) Plots of the particle velocity $u_{\rm o,x}$ versus $t$ following a head-on collisions 
between two heavy particles ($\mathcal{M}=7.5$) for three different 
values of the incident kinetic energy $E_{\rm o}$, at $r_{\rm SR}=1.5\,\xi$.
(c) Plot of the coefficient of restitution $e$ (Eq.~(\ref{eq:cor})) versus 
$E_{\rm o}$, for the head-on collision between two heavy particles ($\mathcal{M}=7.5$).
The inset shows $e$ versus $r_{\rm SR}/\xi$, but for two neutral particles
($\mathcal{M}=1$).}
\label{fig:fig2}
\end{figure*}

We first examine a head-on, two-particle collision. 
We prepare an initial state with two neutral particles, at rest,
separated by $r_0=7\,\xi$ in the superfluid
\footnote{\label{note:ITP} We
prepare such a state by specifying the locations of the particle
$q_{i}$ in Eq.~(\ref{eq:GPEfull}), with $t$ replaced by $-it$,
and integrate it to obtain the ground state. This procedure
yields states whose initial temporal evolution, in the GPE,
produces minimal sound emission~\cite{mfcoeff,nore1997}.}.
We evolve this state by using the 
GPE in the
presence of the SR repulsion between the particles, with $r_{\rm
SR}=1.5\,\xi$, after they are released from rest at $t=0$.
In Fig.~\ref{fig:fig1}(a) we plot the particle positions versus the scaled time $ct/\xi$.
In the insets of Fig.~\ref{fig:fig1}(a), we show pseudocolor plots of
$\rho(\mathbf{r})$ at times labeled ($i$)-($vi$);
these plots show sound waves after the collision
between the particles, which appear as blue disks with
$\rho=0$. We see that the particles
accelerate towards each other and stop on collision,
when the separation $r\simeq r_{\rm SR}$; and then
their motion is reversed, but they do not escape to infinity and undergo
multiple collisions, which are accompanied by acoustic emission,
until they lose their initial kinetic
energy and they stick to form a bound pair; i.e., we have an
inelastic collision (see the spatiotemporal evolution in Video M1, 
Supplemental Material~\cite{suppmat}).

To characterize the SM attractive potential between the
particles, we write the total energy contained in the superfluid
field as
\begin{equation}\label{eq:fieldE}
\begin{split}
	E_{\rm F} &= \frac{1}{\mathcal{A}}\int_{\mathcal{A}}\biggl[\frac{\hbar^2}{2m}|\nabla\psi|^2 + \frac{g}{2}(|\psi|^2-\frac{\mu}{g})^2\\
	   &+\sum_{\rm i=1}^{\mathcal{N}_{\rm o}}V_{\mathcal{P}}(\mathbf{r}-\mathbf{q}_{i})|\psi|^2\biggr]d\mathbf{r}.
\end{split}
\end{equation}

We now perform DNSs, in which we vary the initial
scaled distance $r/\xi$ between the particles; we then obtain
$E_{\rm F}(r/\xi)$, the energy of the minimum-energy state without
the SR repulsion, by using the
imaginary-time procedure~\footnotemark[\value{footnote}].
In Fig.~\ref{fig:fig1}
(b) we plot versus $r/\xi$ the potential $U_{\rm A} = E_{\rm
F}(r)-E_{\rm F}(r=\infty)$, which is negative (i.e., attractive),
for small $r/\xi$, and vanishes in the limit
$r/\xi\rightarrow\infty$. 

We can estimate $U_{\rm A}(r/\xi)$, for the two-particle case, by
using the Thomas-Fermi (TF) approximation~\cite{Pethickbook2001} as follows. We neglect
the kinetic-energy term in Eq.~(\ref{eq:GPEfull}) and write
\begin{equation}
|\psi(\mathbf{r})|^2 = (\mu-\mathcal{V}_{\rm P})\theta(\mu-
\mathcal{V}_{\rm P})/g,
\end{equation}
with $\mathcal{V}_{\rm P}
=V_{\mathcal{P}}(\mathbf{r}-\mathbf{q}_{1})+
V_{\mathcal{P}}(\mathbf{r}-\mathbf{q}_{2})$ and $\theta$ the
Heaviside function that ensures $|\psi|^2>0$.  In this
approximation, 
\begin{equation}\label{eq:potTF}
	E^{\rm TF}_{\rm F} = \frac{1}{\mathcal{A}}\int_{\mathcal{A}}\biggl[\mu^2-(\mu
-\mathcal{V}_{\rm P})^2\theta(\mu-\mathcal{V}_{\rm P})\biggr]/(2g)d\mathbf{r};
\end{equation}
$U^{\rm TF}_{\rm A}=E^{\rm TF}_{\rm F}(r)-E^{\rm TF}_{\rm
F}(r=\infty)$, which we plot in the inset of
Fig.~\ref{fig:fig1}~(b) versus $r/\xi$. It is in qualitative agreement with $U_{\rm A}$ from our DNS;
the quantitative difference arises because the
TF approximation neglects the kinetic-energy term in
Eq.~(\ref{eq:GPEfull}). 

\begin{figure*}
\centering
\includegraphics[width=0.45\linewidth]{3a.eps}
\put(-195.,155){\large{\bf (a)}}
\hspace{0.1cm}
\includegraphics[width=0.45\linewidth]{3b.eps}
\put(-200,155){\large{\bf (b)}}
\put(-325,113){
\includegraphics[width=0.15\linewidth]{3a1.eps}}
\put(-180,125){
\includegraphics[width=0.08\linewidth]{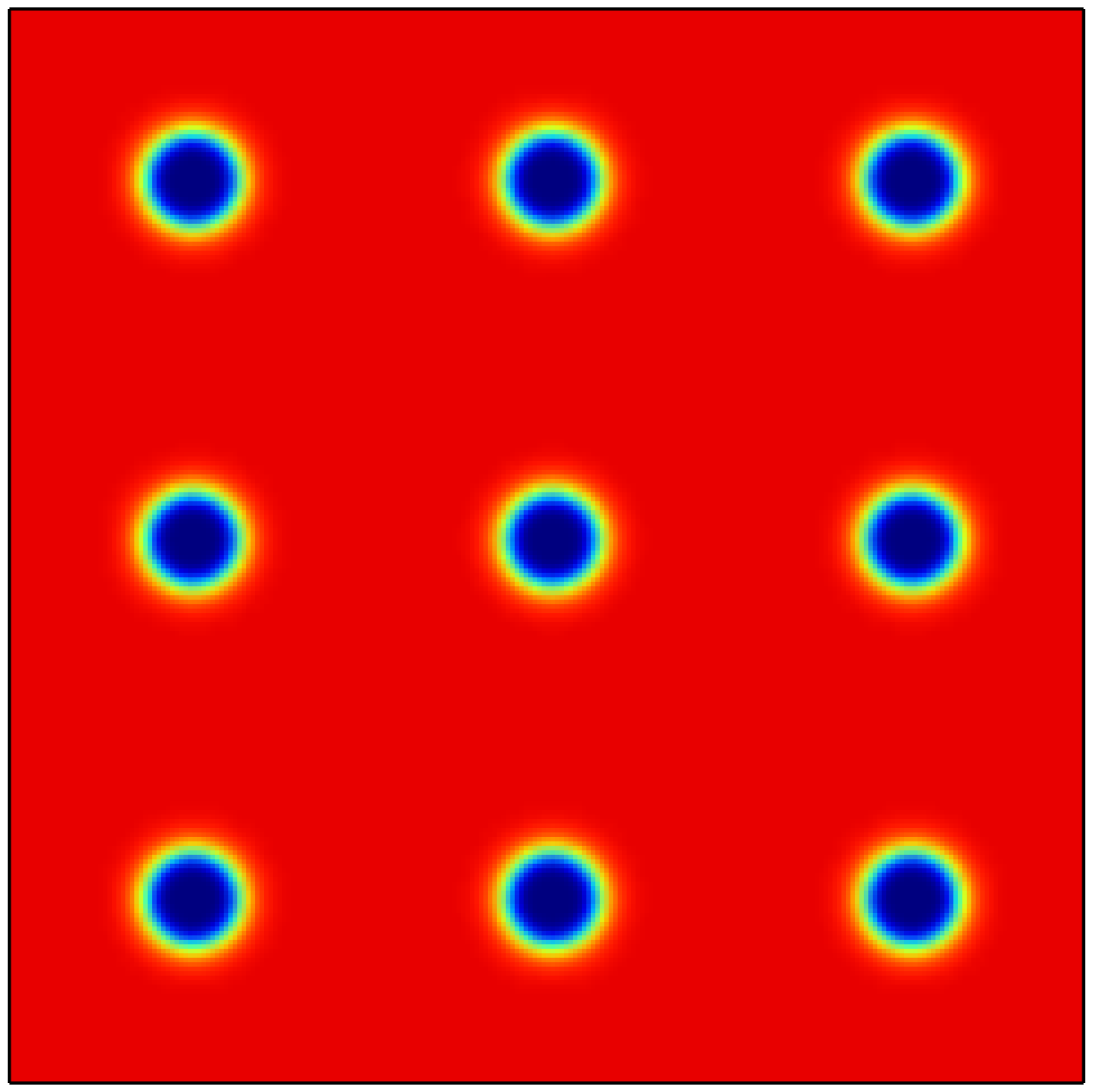}
\put(-40,25){\small{\bf\color{white} (i)}}}
\put(-140,125){
\includegraphics[width=0.08\linewidth]{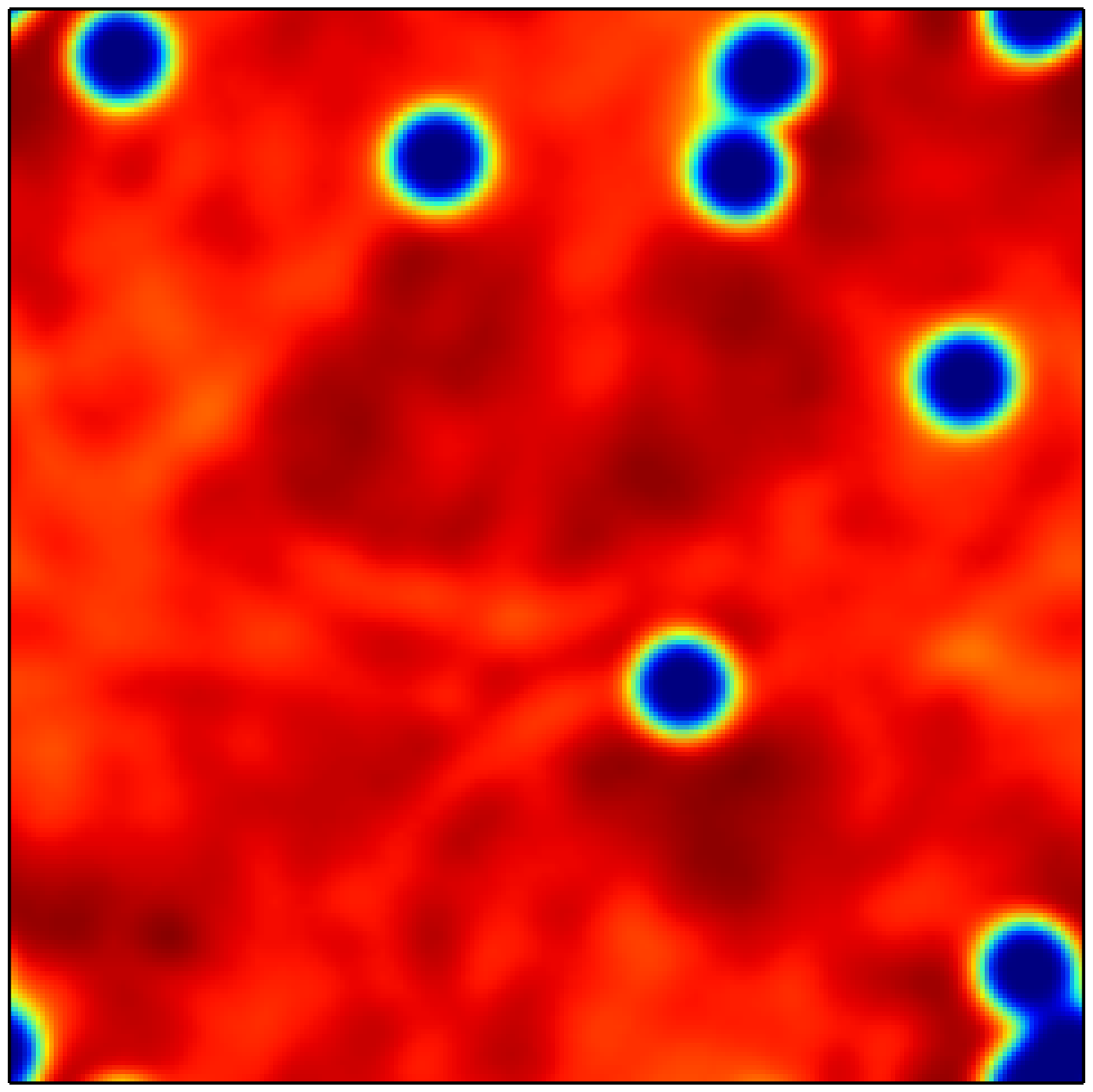}
\put(-40,25){\small{\bf\color{white} (ii)}}}
\put(-100,125){
\includegraphics[width=0.08\linewidth]{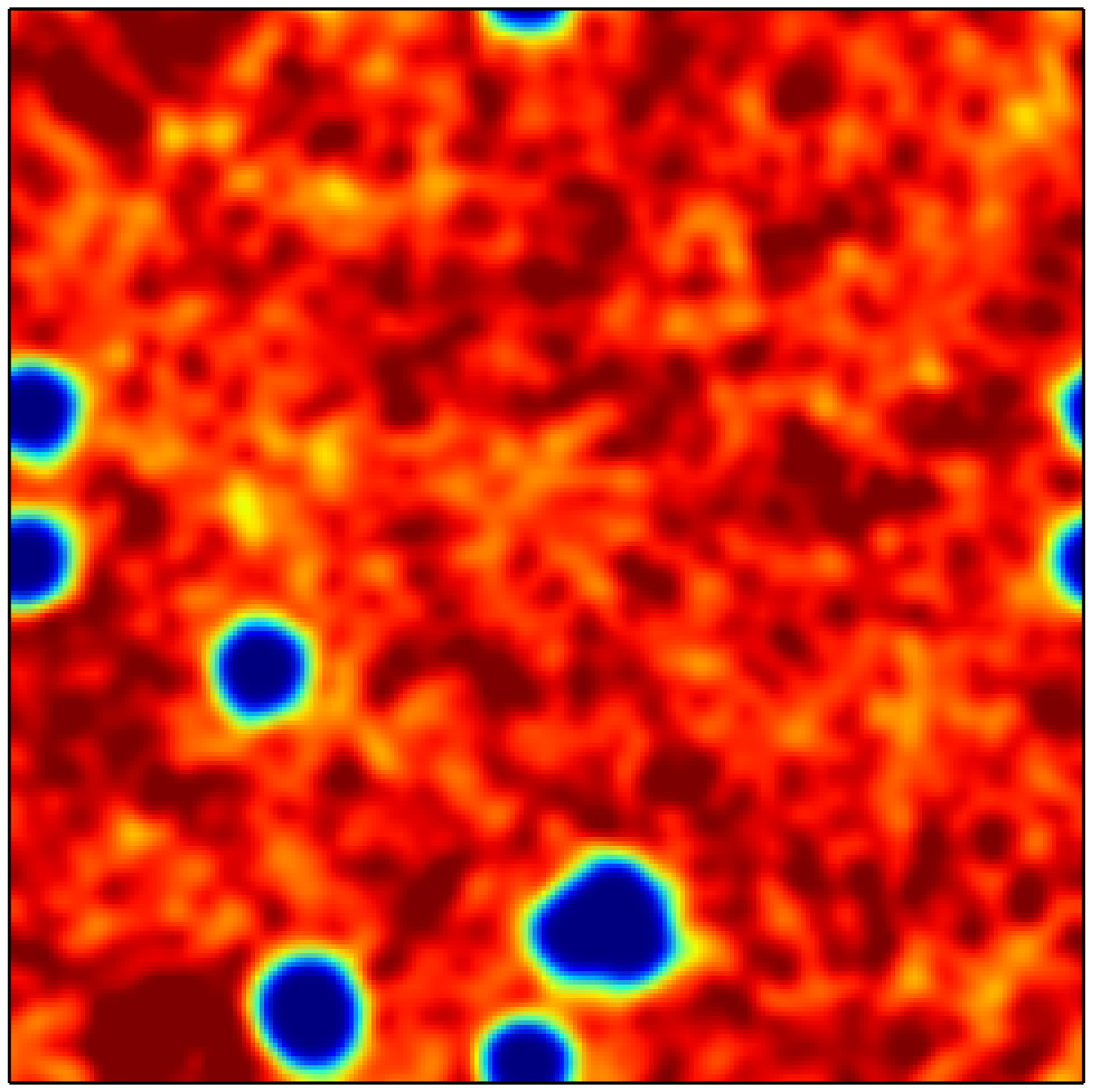}
\put(-20,25){\small{\bf\color{white} (iii)}}}
\put(-60,125){
\includegraphics[width=0.08\linewidth]{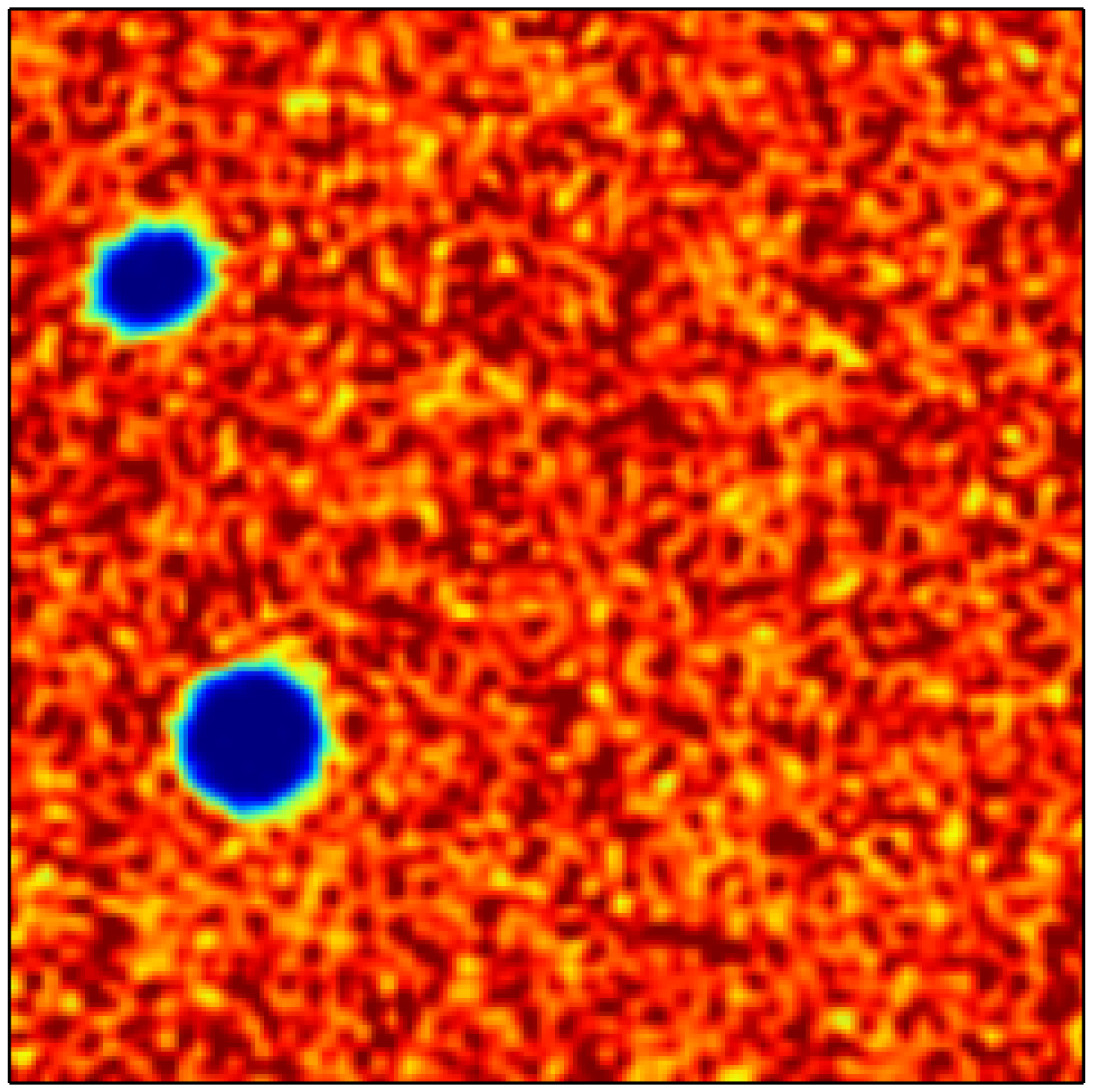}
\put(-20,25){\small{\bf\color{white} (iv)}}}
\caption{\small (Color online)  
	(a) {\bf Collisions at impact parameters $b\geq0$:} 
Light green, dark green, and  blue curves show the particle trajectories
for two heavy particles ($\mathcal{M}=7.5$) $O_{\rm 1}$ (solid curves) 
and $O_{\rm 2}$ (dashed curves) colliding at $b=0$, $b=4\,\xi$, and
$b=2\,\xi$, respectively, with incident kinetic energy $E_{\rm o}\simeq8.7\times 10^{-4}$. For 
$b=2\,\xi$ and $E_{\rm o}=9.1\times10^{-6}$, the colliding particles stick to form
a bound pair (red curves); the inset shows an enlarged view of the particle trajectories 
for the bound-pair, the particle moiton is a quasi-periodic function of time.
(b) {\bf Aggregation:} Plots of the time evolution of $E_{\rm F}(t)-E_{\rm F}(t_0)$, $E_{\rm o}$,
and $E_{\rm SR}$ for nine heavy particles
($\mathcal{M}=7.5$) initially placed on a lattice; these are set into motion by the application of
constant-in-time forces, random in magnitude and direction, for a short duration $t\leq t_{\rm 0}\sim 85$.
The insets (i)-(iv) illustrate multi-particle collisional dynamics, at the representative times
$t_{(i)}=0 < t_{(ii)} < t_{(iii)} < t_{(iv)}$
by pseudocolor plots of the density field $\rho(\mathbf{r})$;
the particles appear as blue disks in which $\rho=0$.
}
\label{fig:fig3}
\end{figure*}

We now study two simplified cases: (1) head-on
collisions, with impact parameter $b=0$; (2) and collisions with
finite, but small $b$. The schematic diagram in
Fig.~\ref{fig:fig2} (a) outlines our procedure. We use an initial
state with two stationary particles $O_{\rm 1}$, at
$(\pi/2,\pi)$, and $O_{\rm 2}$, at $(3\pi/2,\pi)$. We apply the external forces $F_{\rm
ext}=F_{\rm 0}\hat{\mathbf{x}}$ and $F_{\rm ext}=-F_{\rm
0}\hat{\mathbf{x}}$, respectively, to accelerate the
particles; and then we turn off $F_{\rm ext}$ at $t=t_{0}$ (red
vertical line in Fig.~\ref{fig:fig2} (a)).  In
Fig.~\ref{fig:fig2} (b) we plot  versus $ct/\xi$ the $x$ components of the 
particle velocities $u_{\rm o, x}(t)$, 
from our DNS with two heavy particles ($\mathcal{M}=7.5$ and $r_{\rm SR}=1.5\,\xi$), for
three different values of the incident kinetic energy $E_{\rm
o}$.  For $E_{\rm o}=1.5\times 10^{-4}$ (blue curves in
Fig.~\ref{fig:fig2} (b)), the behavior of $u_{o,x}(t)$ is similar
to that of neutral-particle collisions with SR repulsion 
(Fig.~\ref{fig:fig1} (a)); the collision is completely inelastic and
the particles form a bound pair; and the
separation between their centers fluctuates around $r\simeq
r_{\rm SR}$.  The time average of the velocities of the particles
is zero, after the collision.  Figure~\ref{fig:fig2} (b)
shows that, for $E_{\rm o}=1.8\times 10^{-4}$ (green
curves), the two particles rebound,
with small non-zero, mean velocities; at the time of the
collision, most of the energy is transferred to the repulsive
term because of the change in $E_{\rm F}(t)-E_{\rm F}(t_0)$ and $E_{\rm o}$ (see
the Supplemental Material~\cite{suppmat}).  After the collision,
most of the energy is transferred back to the fluid and the
particles have a small kinetic energy. For higher values,
e.g., $E_{\rm o}=5.04\times 10^{-4}$ (magenta curves), the head-on collision
between the heavy particles is nearly
elastic; and the particles rebound with velocities that are
significant fractions of their values at incidence
(see the spatiotemporal evolution in Videos
M2, M3 and M4 in the Supplemental Material~\cite{suppmat}).

We characterize this inelastic-elastic transition by calculating the 
coefficient of restitution for head-on collisions:
\begin{equation}\label{eq:cor}
e = \frac{u_{\rm 2,F}-u_{\rm 1,F}}{u_{\rm 1,I}-u_{\rm 2,I}},
\end{equation}
where $u_{\rm 1,I}$ and $u_{\rm 2,I}$ are, respectively, the mean
velocities of the particles $O_{\rm 1}$ and $O_{\rm 2}$ before
the collision and $u_{\rm 1,F}$ and $u_{\rm 2,F}$ are the mean
velocities of these particles after the collision.  
For the
collisions described above, we find: (1) $e\simeq 0$ for $E_{\rm
o}=1.5\times 10^{-4}$; (2) $e\simeq 0.24$ for $E_{\rm
o}=1.8\times 10^{-4}$; and (3) $e\simeq 0.68$ for $E_{\rm
o}=5.0\times 10^{-4}$.  In Fig.~\ref{fig:fig2} (c) we plot $e$
versus $E_{\rm o}$ from the two-particle, head-on collisions. 
At low values of
$E_{\rm o}$, the particle collision is inelastic with $e=0$; and,
as we increase $E_{\rm o}$, $e$ becomes finite at a critical
value $E_{\rm o}\simeq 1.6\times 10^{-4}$, and then there is a
steep increase followed by a slow, asymptotic growth towards a
value close to $1$.  We observe a similar inelastic-elastic transition, when instead of
$E_{\rm o}$, we vary $r_{\rm SR}/\xi$;
here we consider neutral particles ($\mathcal{M}=1$) to illustrate 
that the sticking transition does not necessarily require 
heavy particles. The plot of $e$ versus $r_{\rm
SR}/\xi$ in the inset of Fig.~\ref{fig:fig2} (c), shows that, at
low values of $r_{\rm SR}/\xi$, the particle collision is
inelastic with $e=0$; and, as we increase $r_{\rm SR}/\xi$, $e$
becomes finite at a critical value $r_{\rm SR}/\xi\simeq 6.46$, and
finally attains a value close to $1$.  

Given the resolution of our study, our data are consistent with a
\textit{continuous sticking transition} at which $e$ goes
to zero continuously as a power $\beta$ of the control parameter
(either $E_{\rm o}$ or $r_{\rm SR}/\xi$). We now give a mean-field
calculation of this power-law exponent $\beta$. The symmetry of these head-on collisions
allows us to write $u_{\rm I}\simeq -u_{\rm 1,I}\simeq u_{\rm 2,I}$
and $u_{\rm F}\simeq u_{\rm 1,F}\simeq -u_{\rm 2,F}$.  The energy
balance between the states, before and after the collision, is
$E_{\rm rad}(u_I)+m_o u_F^2=m_o u_I^2$, where $E_{\rm rad}$ is
the energy radiated into sound waves.  Therefore,
\begin{equation}
e(u_I)=\sqrt{1- E_{\rm rad}(u_I)/m_o u_I^2}, 
\end{equation} 
which yields the critical velocity $u_{\rm I}^c$ at which
$e(u_{\rm I}^c)$ first becomes nonzero.  In a simple, mean-field
approximation, the Taylor expansion of $E_{\rm rad}(u_I)$, around
$u_I=u_I^c$, yields the mean-field (MF) exponent $\beta^{{\rm
MF}}=1/2$. Our DNSs yield values of $\beta$ that are comparable
to, but different from, $\beta^{{\rm MF}}=1/2$. The calculation of
$\beta$ for this sticking transition, beyond our
mean-field theory, and its universality, if any, is a challenging
problem.

In Fig.~\ref{fig:fig3} (a) we show the trajectories of two heavy
particles ($\mathcal{M}=7.5$ and $r_{\rm SR}=1.5\,\xi$) that collide with each other,
with an impact parameter $b > 0$. If the incident kinetic energy
of the particles is sufficiently high, e.g., $E_{\rm o}\simeq
8.7\times 10^{-4}$, they do not stick; the
particles get deflected from their incident trajectory at an
angle $\Theta$, which depends on $b$ (see  Fig.~\ref{fig:fig3}
(a) for $b=2\,\xi$ and $b=4\,\xi$).  However, for $b=2\,\xi$ with
$E_{\rm o}\simeq 9.0\times 10^{-6}$, the incident kinetic energy is
small enough to allow the formation of a bound pair (red curves
in Fig.~\ref{fig:fig3} (a)); the inset shows an enlarged version
of the particle trajectories, after the collision, with red
solid (dashed) curves for particle $O_{\rm 1}$ ($O_{\rm 2}$). The \textit{sun-flower-petal}
pattern of these trajectories indicates that, after transients
have decayed, the damped, oscillatory motion of the particles in
the bound-pair is akin to that of a dimer, with
vibrational and rotational degrees of freedom.
The power-spectra of the time series $q_{i,j}(t)$, for particle
$i\in{\{1,2\}}$ and coordinate $j\in{\{x,y\}}$, show three prominent
frequencies, $\omega_a=0.0185c/\xi$, $\omega_b=0.0148c/\xi$, and
$\omega_c=0.0222c/\xi$, with $2\omega_a=\omega_b+\omega_c$, i.e.,
the oscillatory motion is quasiperiodic (data not shown).  

If we start with more than two particles, then a succession of
inelastic collisions can lead to the formation of multi-particle
aggregates. We illustrate this in Fig.~\ref{fig:fig3}(b) for 
an assembly of $9$ particles
($\mathcal{M}=7.5$ and $r_{\rm SR}=1.5\,\xi$); to initialize the system, we place the
particles on a lattice (inset (i)) and set them into motion by
applying constant-in-time forces, with random magnitudes and
directions, for a given duration, such that the collisions occur
only after the forces are switched off at $t=t_0$. In
Fig.~\ref{fig:fig3}(b) we plot $E_{\rm F}(t) - E_{\rm F}(t_0)$,
$E_0$, and $E_{\rm SR}$ versus $ct/\xi$; large spikes in these plots
occur at collisions; subsequent rearrangements into clusters give
rise to strong fluctuations in $E_{\rm SR}$; as the clusters
settle into their optimal configurations, the fluctuations in
$E_{\rm SR}$ decrease until they saturate towards the end of
our DNS.  The pseudocolor plots of $\rho(\mathbf{r})$ in the insets (ii)-(iv)
of Fig.~\ref{fig:fig3}(b) show the aggregation of particles and
the formation of a $7$-particle cluster and a dimer
(see the spatiotemporal evolution in Video M5 in the Supplemental Material~\cite{suppmat});
a much longer DNS should lead to a $9$-particle cluster here.

In conclusion, our minimal model of active and interacting particles in the
Gross-Pitaevskii superfluid yields qualitatively new results,
like the sticking transition and rich aggregation dynamics of
particle assemblies. Our qualitative results should hold even in
superfluids like Helium, in BECs~\cite{berloff2014modeling}, and in three dimensions.  
Particles in superfluids have been considered by using
Biot-Savart methods~\cite{kivotides08,kivotidePhysFluid08,minedaveldistro,Varga2015}, 
a two-fluid model~\cite{tracerPoole2005};
the GPE has been studied with a single spherical particle~\cite{winiecki1},
however, these studies have not considered the collisions and aggregation we elucidate.
Impurities in BECs~\cite{Klein2007} have been described in
terms of generalized Bose-Hubbard Models~\cite{Spethmann2012},
but these works do not study the problems we consider. We hope
our work will lead to experimental studies of particle collisions
and aggregation in superfluids and BECs.

\acknowledgements
We thank the Council of Scientific and Industrial Research (India),
University Grants Commission (India), Department of Science and
Technology (India) and the Indo-French Centre for Applied Mathematics
for financial support, and Supercomputing
Education and Research Centre, IISc, India for computational
resources. V.S. acknowledges support from Centre Franco-Indien
pour la Promotion de la Recherche Avanc\'ee (CEFIPRA) Project No.
4904.  VS and RP thank ENS, Paris for hospitality and MB
thanks IISc, Bangalore for hospitality.

\bibliographystyle{apsrev4-1}
\bibliography{2dgpepartcoll_vmr}

\setcounter{equation}{0}
\clearpage
\vspace{0.5cm}
{\large\bf{SUPPLEMENTAL MATERIAL}}
\vspace{0.5cm}

\section{Conserved Quantities}
We treat the superfluid and particles together as a single system, in which we use
the Gross-Pitaevskii equation (GPE) to obtain the spatiotemporal
evolution of the condensate wave function $\psi(\mathbf{r},t)$:
\begin{equation}\label{SMeq:GPEful}
i\hbar\frac{\partial \psi}{\partial t} = -\frac{\hbar^2}{2m}\nabla^2\psi -\mu\psi + g|\psi|^2\psi
+ \sum^{\mathcal{N}_0}_{i=1}V_{\mathcal{P}}(\mathbf{r}-\mathbf{q}_i)\psi;
\end{equation}
here $g$ is the effective interaction strength, $m$ the mass of the bosons, $\mu$ the 
chemical potential, $V_{\mathcal{P}}$ the potential that we use to represent the 
particles, and $\mathcal{N}_0$ the total number of particles with mass $m_{o}$.
The equation of motion for the particle $i$ is
\begin{equation}\label{SMeq:eqmpartfull}
	m_{o}\ddot{\mathbf{q}}_i = \mathbf{f}_{o,i} + \mathbf{f}_{SR,i} + \mathbf{F}_{\rm
ext,i},
\end{equation}
where
\begin{equation}\label{eq:fospart}
	\mathbf{f}_{o,i} = \int_{\mathcal{A}}|\psi|^2\nabla V_{\mathcal{P}}d\mathbf{r};
\end{equation}
$\mathbf{f}_{SR,i}$ arises from the short-range (SR) repulsive potential, 
$\mathbf{F}_{\rm ext,i}$ is the external force acting on the $i$th particle,
$\mathbf{q}_i$ the particle position vector (the overhead dot represents differentiation
with respect to time)
and ${\mathcal{A}}$ is the simulation
domain.

In the 
absence of any external applied force, the total energy $E$ of this system is conserved. 
We write the total energy of the system as 
\begin{equation}
E = E_{\rm F} + E_{\rm o} + E_{\rm SR};
\end{equation}
here $E_{\rm F}$ is the energy contained in the superfluid field, $E_{\rm o}$ the total kinetic energy
of the particles, and $E_{\rm SR}$ the energy from the SR repulsion between the particles;
these energies are defined, respectively, as follows:
\begin{subequations}
\begin{align}
	E_{\rm F} &= \frac{1}{\mathcal{A}}\int_{\mathcal{A}}\biggl[\frac{\hbar^2}{2m}|\nabla\psi|^2 
	+ \frac{1}{2}g\Bigl(|\psi|^2-\frac{\mu}{g}\Bigr)^2\\
&+\sum\nolimits_{\rm i=1}^{\mathcal{N}_{\rm o}}
V_{\mathcal{P}}(\mathbf{r}-\mathbf{q}_{i})|\psi|^2\biggr]d^2x; \\
E_{\rm o} &= \frac{1}{\mathcal{A}}\sum\nolimits_{\rm i=1}\frac{1}{2}m_{\rm o}\dot{\mathbf{q}}_{i}^2;\\
E_{\rm SR} &= \frac{1}{\mathcal{A}}\sum\nolimits_{\rm i,j, i\neq j}^
{\mathcal{N}_{\rm o},\mathcal{N}_{\rm o}} \frac{\Delta_E r^{12}_{\rm SR}}{|q_{i}-q_{j}|^{12}};
\end{align}
\end{subequations}
here $\mathcal{N}_{\rm o}$ is the number of particles and $\mathcal{A}$ the area of our two-dimensional, computational domain.

The dynamical evolution of the GPE coupled with the equation 
of motion for the particles Eq.~(\ref{SMeq:eqmpartfull})-(\ref{eq:fospart})
conserves the total momentum 
\begin{equation}
\begin{split}
\mathbf{P}(t) &=\mathbf{P}(t=0) + \mathbf{F}_{\rm ext}t  \\
		     &=  \int_{\mathcal{A}}\frac{i\hbar}{2}(\psi^*\nabla\psi-\psi\nabla\psi^*)d^2x
+ \sum\nolimits_{\rm i=1}^{\mathcal{N}_{\rm o}} m_{\rm o}\dot{\mathbf{q}}_{i} \\
&+ \mathbf{F}_{\rm ext}t;
\end{split}
\end{equation}
and  the total number $N$ of bosons, constituting the superfluid,
\begin{equation}
N =\int_{\mathcal{A}}|\psi|^2d^2x.
\end{equation}

\section{Pseudospectral Method and time-stepping}
We use a Fourier pseudospectral method to solve the GPE numerically. 
Thus the fields in spectral space are obtained through a discrete 
Fourier transform with a finite number of modes. We 
introduce the Galerkin projector $\mathcal{P}_G$
\begin{equation}
\mathcal{P}_G[\hat{\psi}(k)]= \theta(k_{\rm max}-k)\hat{\psi} (k),
\end{equation}
where $\hat{\psi}$ is the spatial Fourier transform of $\psi$, 
$k_{\rm max}$ is a suitably chosen ultra-violet cut-off, and
$\theta(\cdot)$ the Heaviside function.
To ensure global momentum conservation we use the
$2/3$-dealiasing rule, with $k_{\rm max}=2/3\times N_c/2$,
where $N_c$ is the number of collocation points~\cite{giorgio2011longPRE}. This conservation of momentum is essential for
the study of the collision between particles. Thus, the Galerkin-truncated GPE
(henceforth TGPE) is
\begin{equation}\label{eq:tgpe} 
\begin{split}
i\hbar\frac{\partial
\psi(\mathbf{r},t)}{\partial t} &= \mathcal{P}_G\Biggl[
	\Bigl(-\frac{\hbar^2}{2m}\nabla^2 + g\mathcal{P}_G[|\psi|^2] \\
&-\mu
+ \sum\nolimits_{\rm i=1}^{\mathcal{N}_{\rm o}}
V_{\mathcal{P}}(\mathbf{r}-\mathbf{q}_{i})\Bigr)\psi(\mathbf{r},t)
\Biggr].
\end{split}
\end{equation}
In our Galerkin-truncation scheme, we can write Eq.~(\ref{eq:fospart}) as
\begin{equation}
\begin{split}
\mathbf{f}_{{\rm o},i} &= -\int_{\mathcal{A}}\biggl[\psi^*\mathcal{P}_G
[V_{\mathcal{P}}(\mathbf{r}-\mathbf{q}_{i})\nabla\psi] \\
& + \psi\mathcal{P}_G[V_{\mathcal{P}}(\mathbf{r}-\mathbf{q}_{i})\nabla\psi^*]\biggr]d^2x.
\end{split}
\end{equation}

To perform a direct numerical simulation (DNS) of the TGP Eq.~(\ref{eq:tgpe}),
together with Eq.~(\ref{SMeq:eqmpartfull}) that is coupled with it, we have developed a parallel, 
MPI code in which we discretize $\psi(\mathbf{r},t)$ on a square, periodic simulation domain of side $L=2\pi$ with
$N^2_c=128^2$ collocation points.
We evaluate the linear terms in Eq.~(\ref{eq:tgpe}) in Fourier space and
the nonlinear term in physical space; we use the FFTW library~\cite{fftw}
for the Fourier-transform operations, and
a fourth-order, Runge-Kutta scheme, with time step 
$\Delta t$, to evolve equations Eqs.~(\ref{SMeq:eqmpartfull}) and~(\ref{eq:tgpe}) in
time.

For an assembly of 9 particles (for a full discussion see the Main text), the energy and momentum are
conserved upto $2\times 10^{-5}$ and $2\times 10^{-7}$ percent, respectively;
our DNS uses $5\times 10^6$ steps with $\Delta t=2\times 10^{-4}$, and $N_c=128$.

\section{Supplemental figures}

\begin{figure*}[!htbp]
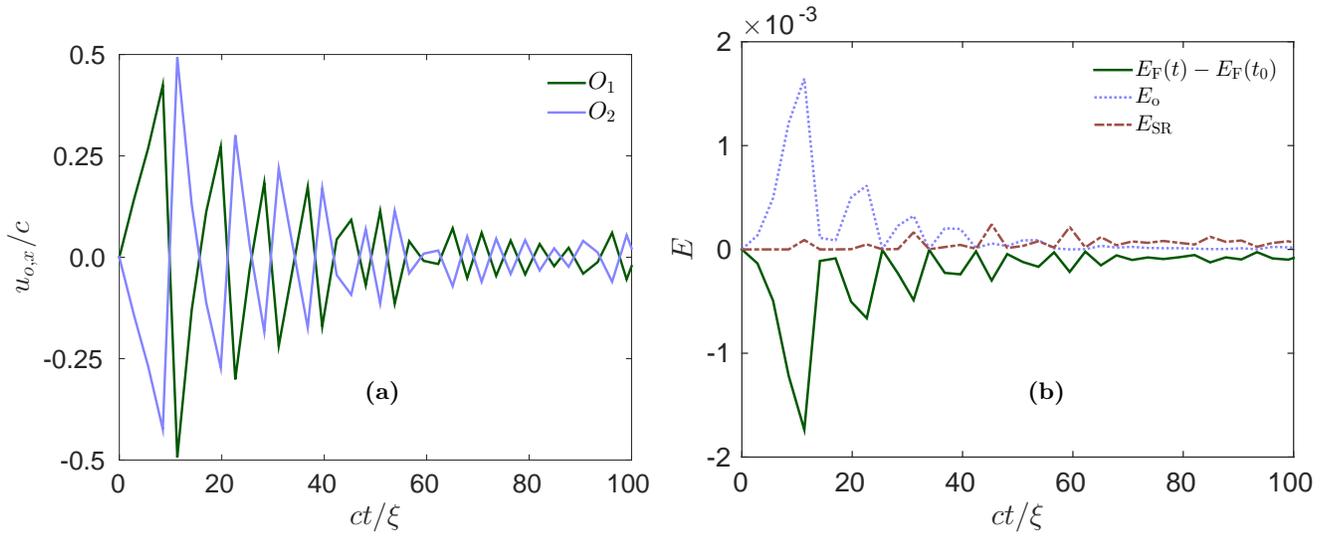

\centering
\includegraphics[width=0.48\linewidth]{S1b.eps}
\put(-110.,50){\small{\bf (a)}}
\hspace{0.1cm}
\includegraphics[width=0.48\linewidth]{S1c.eps}
\put(-110.,50){\small{\bf (b)}}
\caption{\small (Color online) {\bf Superfluid-mediated attractive potential:}
	(a) Plots of the particle velocity $u_{\rm o,x}$ 
	versus the scaled time $ct/\xi$;
	(b) plots of the time evolution of the energies $E_{\rm F}(t)-E_{\rm F}(t_0=0)$ (green solid curve), 
	$E_{\rm o}$ (blue dotted curve), and $E_{\rm SR}$ (magenta dashed curve),
	following a head-on collision between two neutral particles 
	($\mathcal{M}=1$ and $r_{\rm SR}=1.5\,\xi$). The particles are released from rest with an initial
	separation $r_0=7\,\xi$. (For a full discussion see the Main text.)
}
\label{fig:figS1}
\end{figure*}

\subsection{{\textcolor{blue}{\bf Figure~\ref{fig:figS1}:}} Superfluid-mediated attractive potential}

\begin{figure*}[!htbp]
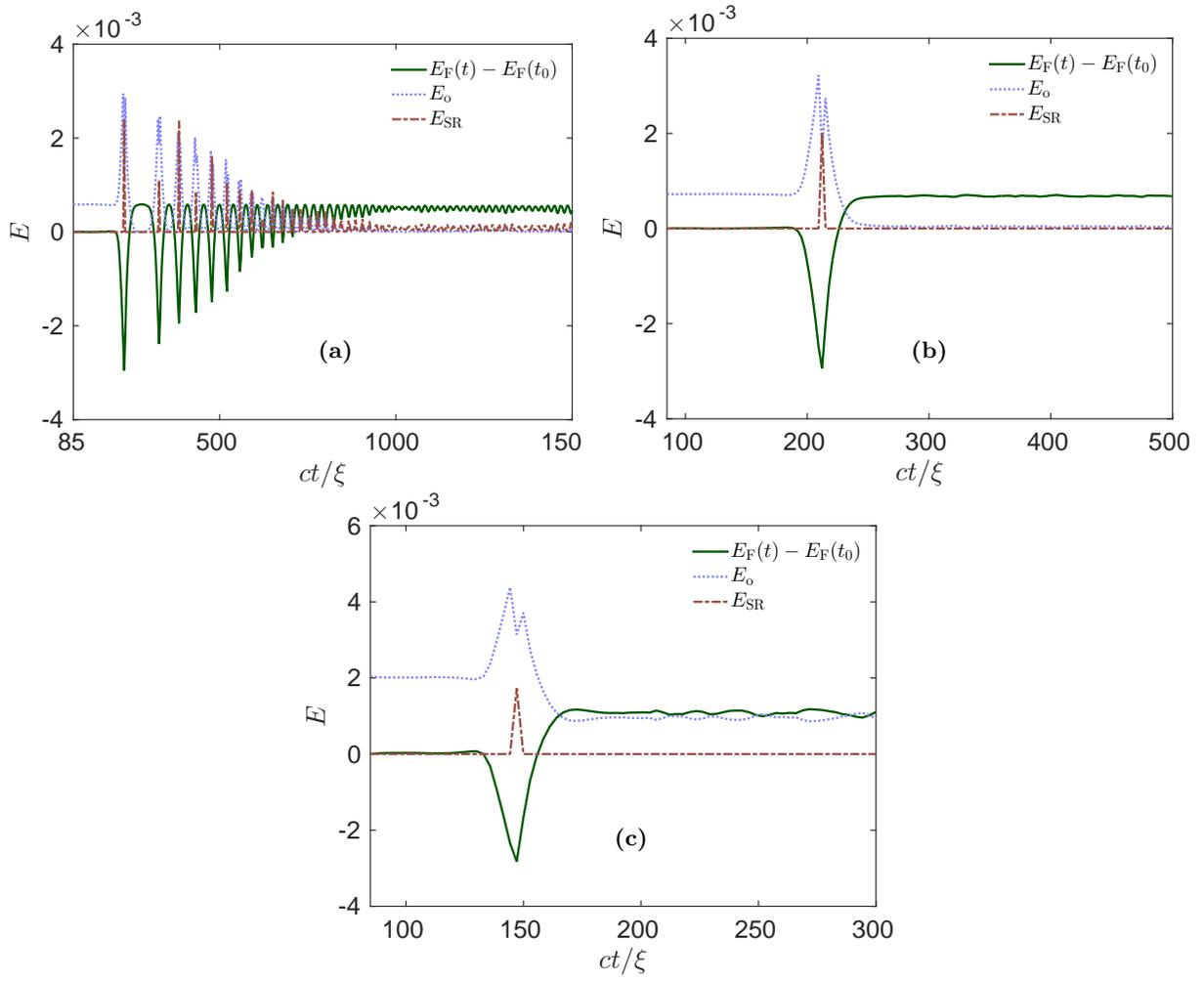

\centering
\includegraphics[width=0.45\linewidth]{S2a.eps}
\put(-110.,50){\small{\bf (a)}}
\includegraphics[width=0.45\linewidth]{S2b.eps}
\put(-110.,50){\small{\bf (b)}}\\
\includegraphics[width=0.45\linewidth]{S2c.eps}
\put(-110.,50){\small{\bf (c)}}
\caption{\small (Color online)  {\bf Head-on collisions:}
	Plots of the time evolution of the energies $E_{\rm F}(t)-E_{\rm F}(t_0=85)$ (green solid curve), 
	$E_{\rm o}$ (blue dotted curve), and $E_{\rm SR}$ (magenta dashed curve)
	for three different values of the incident kinetic energy (a) $E_{\rm o}=1.5\times 10^{-4}$, 
	(b) $E_{\rm o}=1.8\times 10^{-4}$, and $E_{\rm o}=5.0\times 10^{-4}$, following a head-on collision
	between two heavy particles ($\mathcal{M}=7.5$ and $r_{\rm SR}=1.5\,\xi$).
	(For a full discussion see the Main text.)
}
\label{fig:figS2}
\end{figure*}

\subsection{{\textcolor{blue}{\bf Figure~\ref{fig:figS2}:}} Head-on collisions}

\section{Videos}

{\textcolor{blue}{\bf Video M1}}(\url{https://youtu.be/F3mq-raIdTM})
This video illustrates the collisional dynamics of two 
neutral particles ($\mathcal{M}=1$) in the presence of a
short-range, repulsive interaction ($r_{\rm SR}=1.5\,\xi$), 
when they are released from rest. The dynamics of the particles is illustrated by the spatiotemporal 
evolution of the field $\rho(\mathbf{r})$, shown via sequence of pseudocolor plots,
separated by $t=2.83$ (we use $10$ frames per second).

{\textcolor{blue}{\bf Video M2}}(\url{https://youtu.be/2xIUckXILN0})
This video illustrates the collisional dynamics of two heavy particles ($\mathcal{M}=7.5$) for the 
incident kinetic energy $E_{\rm o}=1.5\times 10^{-4}$ and $r_{\rm SR}=1.5\xi$.
The dynamics of the particles is illustrated by 
the spatiotemporal evolution of the field $\rho(\mathbf{r})$, 
shown via sequence of pseudocolor plots,
separated by $t=2.83$ (we use $10$ frames per second).

{\textcolor{blue}{\bf Video M3}}(\url{https://youtu.be/X5356R4_XwM})
This video illustrates the collisional dynamics of two heavy particles ($\mathcal{M}=7.5$) for the incident 
kinetic energy $E_{\rm o}=1.8\times 10^{-4}$ and $r_{\rm SR}=1.5\xi$.
The dynamics of the particles is illustrated by 
the spatiotemporal evolution of the field $\rho(\mathbf{r})$, 
shown via sequence of pseudocolor plots,
separated by $t=2.83$ (we use $10$ frames per second).

{\textcolor{blue}{\bf Video M4}}(\url{https://youtu.be/fq5Hb0PALAs})
This video illustrates the collisional dynamics of two heavy particles ($\mathcal{M}=7.5$) for the
incident kinetic energy $E_{\rm o}=5.0\times 10^{-4}$ and $r_{\rm SR}=1.5\xi$.
The dynamics of the particles is illustrated by 
the spatiotemporal evolution of the field $\rho(\mathbf{r})$, 
shown via sequence of pseudocolor plots,
separated by $t=2.83$ (we use $10$ frames per second).

{\textcolor{blue}{\bf Video M5}}(\url{https://youtu.be/zqh_3nq-2YY})
This video illustrates the collisional dynamics of nine heavy particles 
($\mathcal{M}=7.5$ and $r_{\rm SR}=1.5\xi$);
initially placed on a lattice and set into motion by the application of 
constant-in-time forces, random in magnitude and direction,
for a short duration $t\simeq85$. The dynamics of the particles 
is illustrated by the spatiotemporal evolution of 
the field $\rho(\mathbf{r})$, 
shown via sequence of pseudocolor plots,
separated by $t=14.15$ (we use $15$ frames per second).


\end{document}